\documentstyle[prl,multicol,psfig,aps]{revtex}
\begin{document}
\draft

\title{Far-infrared reflectivity of Bi$_{2}$Sr$_2$CuO$_{6}$: the 
``anomalous Drude'' model and the optical pseudogap revisited}
\author{S. Lupi, P. Calvani, and M. Capizzi}
\address{Istituto Nazionale di Fisica della Materia - Dipartimento di
Fisica, Universit\`a di Roma La Sapienza, Piazzale Aldo Moro 2, I-00185 
Roma, Italy}
\author{P. Roy}
\address{Laboratoire pour l'Utilisation du Rayonnement
Electromagn\'etique, Universit\'e Paris-Sud, 91405 Orsay, France}
\date{\today}
\maketitle

\begin{abstract}
An optical ``pseudogap'' is assumed to open at low $T$  in the ``anomalous 
Drude'' absorption, which models the optical conductivity $\sigma(\omega) 
\propto \omega^{-1}$ of high-$T_c$ superconductors by a linewidth 
$\Gamma \approx 10^3$ cm$^{-1}$ varying with $\omega$. In the $\sigma(\omega)$ 
of Bi$_{2}$Sr$_2$CuO$_{6}$ measured down to 10 cm$^{-1}$, we have 
resolved, instead, two components separated by a deep minimum: 
i) a normal Drude term with $\Gamma$=35 cm$^{-1}$ 
at 30 K, in very good agreement with transport data; 
ii) a strong band peaked in the far infrared (FIR), 
likely due to bound charges,  whose tail  exhibits the 
$\omega^{-1}$ dependence. As the FIR peak
softens for $T \to 0$, it opens a pseudogap-like depression in 
$\sigma(\omega)$ accordingly to ordinary sum rules.
\end{abstract}
\begin{multicols}{2}
\narrowtext

Several studies have been published during recent years on the possible 
implications for high-$T_c$ superconductivity  of  the  
``pseudogap''. This term indicates a depression in the low-energy continuum 
of states of many high-$T_c$ superconductors (HCTS),
that has been observed
by different techniques  below a characteristic temperature 
$T^*>>T_c$.\cite{Timusk} The pseudogap has been observed also in the
optical spectra of a number of metallic 
cuprates. These include underdoped systems like 
Bi$_{2}$Sr$_2$CuO$_{6}$ with $T_c$ = 8 K\cite{Tsvetkov} and
HgBa$_{2}$Ca$_2$Cu$_3$O$_{8+y}$ with $T_c$ = 121 
K,\cite{McGuire}  overdoped cuprates like 
La$_{2-x}$Sr$_{x}$CuO$_{4}$ with $x > 0.18$\cite{Startseva} 
and Tl$_{2}$Ba$_2$CuO$_{6+y}$,\cite{Puchkov}
and even metals close to optimum doping like  
HgBa$_{2}$Ca$_2$Cu$_3$O$_{8+y}$ with $T_c$ = 130 
K\cite{McGuire} and Bi$_2$Sr$_2$CaCu$_2$O$_{8+y}$ with $T_c$ = 
90 K.\cite{Puchkov96} 
In these experiments, the free-carrier contribution to
the real part $\sigma(\omega)$ of the optical conductivity  is described 
by an ``anomalous Drude'' term\cite{Timusk} 

\begin{equation}
\sigma (\omega) = {\omega_D^2/4\pi \over 
{\Gamma^*(\omega) + [m^*(\omega)/ m]^2\omega^2 /\Gamma^*(\omega)}} 
\, ,
\label{eq1}
\end{equation}

\noindent
where $\omega_D$ is a constant plasma frequency, while both the scattering 
rate $\Gamma^*$ and the effective mass $m^*$ depend
on the photon energy $\omega$. This model has been originally
introduced to fit the $\sigma (\omega) \propto \omega^{-1}$ law which 
replaces 
the $\omega^{-2}$ behavior of conventional metals in the mid-infrared 
spectra of
HCTS. The use of Eq. (1) leads to large scattering rates for the 
carriers in the cuprates. At  $T \sim T^*$, $\Gamma^* \approx 1500$ 
cm$^{-1}$ 
in the far infrared,  for both La$_{2-x}$Sr$_x$CuO$_{4}$ and 
HgBa$_{2}$Ca$_2$Cu$_3$O$_{8+y}$.  Below $T^*$,
$\Gamma^*$ decreases to $\sim$  500 cm$^{-1}$  in the former compound, 
to $\sim$  1000 cm$^{-1}$  in the latter.\cite{McGuire,Startseva} Thus, even 
if the opening of an optical pseudogap 
below $T^*$ is displayed by $\sigma (\omega)$, its amplitude is generally 
extracted from the above drop in the far-infrared (FIR) part of 
$\Gamma^*(\omega)$. The idea of an optical pseudogap is therefore 
intrinsically 
related to the anomalous Drude approach of Eq. (1) and to the assumption that 
the mid-infrared absorption of cuprates is dominated by some anomalous 
Fermi liquid.

However, the above picture leads to some
inconsistencies. For instance, in Bi$_2$Sr$_2$CaCu$_2$O$_{8+y}$ 
samples with similar $T_c$'s, the 
pseudogap determined optically ($\sim 700$ cm$^{-1}$)\cite{Puchkov96} is 
much 
larger than that found in angle-resolved
photoemission (0-260 cm$^{-1}$, depending on the direction in the $k$-
space).\cite{Randeria} This discrepancy can hardly be explained, 
if one also considers that the optical absorption results
from an average on the Fermi surface. Moreover,
an optical pseudogap 
is observed in La$_{2-x}$Sr$_x$CuO$_{4}$,\cite{Startseva} although 
no such effect appears in the spin-lattice relaxation rate of this 
cuprate.\cite{Timusk} Finally, in the metallic spectra of 
Refs. \onlinecite{McGuire,Startseva,Puchkov} $\sigma (\omega)$ 
surprisingly 
decreases for $\omega \to$ 0, for any $T$ of the normal 
phase. This decrease is observed on the low-frequency side of a strong 
FIR peak {\it which is systematically associated with the 
observation of the pseudogap}. However, the FIR peak 
is attributed\cite{McGuire,Startseva}  to scattering of the free carriers by 
oxygen atoms randomly distributed in the Cu-O planes. As this 
explanation does not seem  to be related to the existing theoretical models of 
the pseudogap,\cite{Timusk} the link 
between this latter and the FIR peak remains obscure.  

A first hint comes from the optical conductivity of
Nd$_{1.88}$Ce$_{0.12}$CuO$_4$ (NCCO), 
partially reported in Ref. \onlinecite{prl99}. A peak develops in the
infrared, which softens considerably as $T$ is lowered. This shift 
causes at higher energies a depression in $\sigma (\omega)$ around 
1000 cm$^{-1}$, very similar to the pseudogap reported by the authors cited
above. This suggests that, in NCCO, the optical pseudogap opens by 
a transfer of spectral weight to the FIR peak as this softens 
and narrows, as requested by ordinary sum rules. One should then focus on 
this peak at  finite frequency, attributed in NCCO to charges bound to the 
lattice via a polaronic coupling.\cite{prl99}
However, Nd$_{1.88}$Ce$_{0.12}$CuO$_4$ is electron doped and 
semiconducting, even if very close to the insulator-to-metal transition. 
In order to study in greater detail a system where the 
``anomalous-Drude + pseudogap'' model has already 
been applied\cite{Tsvetkov} and with the lowest carrier density,
we have measured the reflectivity of Bi$_{2}$Sr$_2$CuO$_{6}$ 
(BSCO), from 400 to 8 K and from 15,000 to 10 cm$^{-1}$.

By extending the measurements in BSCO to the very-far-infrared region of 
frequencies, we 
have first resolved from the FIR peak a normal Drude contribution which is in 
agreement with the dc conductivity, $\sigma_{dc}$, and the London 
penetration 
depth of the material. Basing on these results we show that, at least in 
BSCO: i) $\sigma (\omega,T)$ cannot be explained in terms of a 
one-component model like the anomalous Drude of Eq. (1);  ii) the 
pseudogap does not open in the continuum of states of some kind of 
(anomalous) free carrier. On the contrary, we show that: i) only the 
coexistence of two type 
of carriers may account for the optical absorption; ii) the 
optical pseudogap is an effect created by the temperature-dependent 
absorption of some weakly bound charges. In such context,
the phase-separation model proposed by Emery and Kivelson\cite{Emery} is
found to be in good agreement with the present observations.

The sample investigated here is a thick BSCO film, highly 
oriented with the $c$-axis orthogonal to the surface. It has been
grown by liquid phase epitaxy on a LaGaO$_3$ substrate.\cite{Balestrino} 
Its resistivity 
$\rho(T)$, obtained from standard 4-points measurements, is reported 
by a full line in the inset of Fig. 1. It shows a linear dependence on 
$T$ from 300 to about  65 K, where a slight change of slope is observed.
The superconducting transition  has its onset at $T_c$ = 20 K, with a 
width of 5 K. The $\rho$ value at 300 K (1.4 $\times 10^{-3} \Omega$cm) 
is intermediate between that of a good BSCO single crystal 
(0.3 $\times 10^{-3}\Omega$cm)\cite{Martin} and that of a polycrystalline 
pellet of the 
same material (2.7 $\times 10^{-3}\Omega$cm).\cite{Xiao} This may be 
attributed to the
presence of grain boundaries in the well oriented $a-b$ plane of the present 
film. However, eventual grain boundaries are not expected to affect 
the infrared measurements. Indeed, the reflectivity $R(\omega)$ of 
the BSCO film, relative to a gold-plated reference and reported in Fig. 
1, is typical of the best single crystals. The
$R(\omega)$ spectra, with the electric field polarized in the 
$a-b$ plane, were collected using a Bomem DA8 interferometer 
coupled to the infrared synchrotron radiation beamline SIRLOIN of the LURE 
laboratory at Orsay. Helium-cooled bolometers, mercury-cadmium-tellurium 
or silicon detectors were used, depending on the frequency range under 
investigation. The large size of the film (0.5x0.5 cm$^2$) and the good 
performances of the apparatus have allowed us to measure $R(\omega)$ 
down to 
unusually low values of $\omega$ ($\sim$ 10 cm$^{-1}$). The film 
thickness (1.8 $\mu$m) was such that no correction for the substrate 
contribution to $R(\omega)$ was needed; see Fig. 1. Therefore, the optical 
conductivity $\sigma(\omega)$ was obtained by simple, 
canonical Kramers-Kronig
transformations of $R(\omega)$. In the normal metallic phase ($T > T_c$), 
$R(\omega)$ has been extrapolated from $\omega$ = 10 cm$^{-1}$ to 
$\omega$ = 0 by a Drude-Lorentz fit, in the superconducting phase by a 
London conductivity, as usually done; see Ref. \onlinecite{Tanner99}.
Moreover, $R(\omega)$ has been extrapolated from 
15,000 cm$^{-1}$ up to 320,000 cm$^{-1}$ by using the 
data of Ref.\onlinecite{Terasaki}, 
that have been extended to higher frequencies by a $\omega^{-4}$ law.

\begin{figure}
{\hbox{\psfig{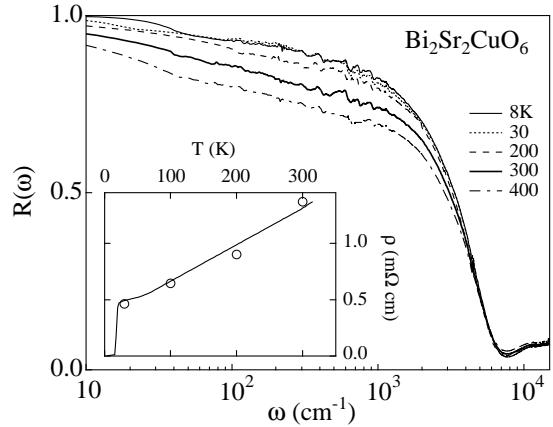}}}
\caption{Reflectivity $R(\omega$) of the BSCO film
at five temperatures from 400 to 8 K. The inset shows the
resistivity $\rho(T)$ of the film from 300 K to 4 K. The open circles are 
the $\rho$  values obtained by  extrapolating $\sigma (\omega,T)$  in Fig. 
3 to $\omega$ = 0.}
\label{1}
\end{figure}

The conductivities $\sigma (\omega,T)$ corresponding to the raw reflectivities 
of Fig. 1 are shown in Fig. 2. At any $T$, $\sigma (\omega,T)$ 
exhibits a broad band 
peaked in the far infrared (FIR peak). Its high-frequency side behaves as 
$\omega^{-1}$ up to 
6000 cm$^{-1}$, as shown in the inset of Fig. 2 for $T$ = 200 and 30 K. 
The tail of the FIR peak accounts, therefore, for the 
frequency dependence of the so-called anomalous-free-carrier absorption 
usually observed in 
metallic cuprates. The peak frequency ($\sim$ 800 cm$^{-1}$ at 400 K) 
increasingly softens and narrows as $T$ is lowered, 
until it reaches 110 cm$^{-1}$ at 30 K.
In the mid infrared, a broad pseudogap opens for $T \to 0$. This feature
is less deep than that observed in other cuprates (see Refs.
\onlinecite{McGuire,Startseva,Puchkov}), most likely because 
the doping of the present BSCO sample is nearly optimal. 
This pseudogap is just due to the red shift and the narrowing of the FIR peak 
for decreasing $T$, which produce a transfer of spectral weight 
towards low frequencies. Indeed, the effective number of 
carriers per unit cell $n_{eff}$ is constant with $T$ within a few percent, 
for all temperatures of the normal phase, see the inset in Fig. 3.
$n_{eff}$ has been evaluated from the relation

\begin{equation}
n_{eff} = {2m^*V\over \pi e^2} \int_{\omega_{min}}^{\omega_{max}} 
\sigma 
(\omega)\, d \omega  \, ,
\label{eq3}
\end{equation}

\noindent
where $V$ is the cell volume, $ \omega_{min}$ = 10 cm$^{-1}$, 
$\omega_{max}$ = 
7500 cm$^{-1}$, and $m^*$ has been assumed equal to the free electron 
mass.

\begin{figure}
{\hbox{\psfig{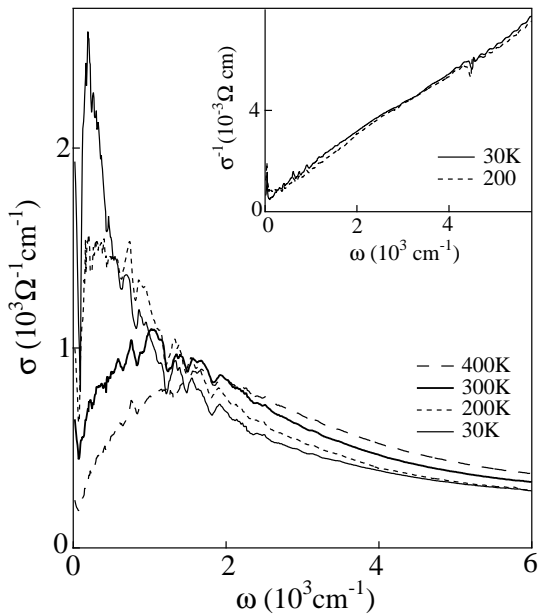}}}
\caption{Optical conductivity $\sigma(\omega)$ of the BSCO film, as 
extracted from $R(\omega$) in Fig. 1 by Kramers-Kronig 
transformations. The inset shows that, on the high-frequency side of the FIR 
peak, $\sigma^{-1}(\omega)$ is linear with $\omega$ both at 200 K 
and 30 K.} 
\label{2}
\end{figure}

The results of Fig. 2 are quite similar to those reported in Ref. 
\onlinecite{Tsvetkov} for a single crystal of BSCO with $T_c$ = 8 K, whose 
$R(\omega)$ was measured from 12,000 to 50 cm$^{-1}$.  However, as 
evident from Fig. 3 where the FIR $\sigma(\omega,T)$ has been reported on 
an expanded scale, the extension of the spectra down to 10 cm$^{-1}$ has 
allowed us to resolve a narrow, low-frequency absorption from the FIR 
peak. The dip separating these two components of $\sigma (\omega)$ is observed
at all temperatures. It corresponds to a change of slope at $\sim$ 50
cm$^{-1}$ in $R (\omega)$, see Fig. 1, where no instrumental effects 
are present, e. g., beamsplitter transmittance minima or data file overlaps.
The low-frequency component of the absorption increases for 
$\omega \to 0$, extrapolates to the $\sigma_{dc}(T)$ values measured 
in the same sample (full triangles), and disappears below $T_c$. 
The open circles in Fig. 3 show that this feature, presumably peaked at 
$\omega =0$, is very well fitted at 30 K by a {\it normal Drude} term

\begin{equation}
\sigma (\omega) =
{\omega_D^2/4\pi \over {\Gamma_D + \omega^2 /\Gamma_D}} \,  ,
\label{eq2}
\end{equation}

\noindent
with  plasma frequency $\omega_D$ = 2100 cm$^{-1}$ and  scattering rate 
$\Gamma_D$ = 35 cm$^{-1}$. At 100 (200) K one obtains, instead, 
$\omega_D$ = 1800 (1600) cm$^{-1}$ and  $\Gamma_D$ = 35 (40) cm$^{-1}$.
If one extrapolates the Drude fits to obtain $\sigma (0,T)$ at all 
temperatures, one obtains the values reported by open circles 
in the inset of Fig. 1 together with the dc resistivity (full line) 
measured in the same sample. The agreement is within a few percent.

The results reported in Figs. 2 and 3 point towards a
multi-component model for $\sigma (\omega)$ and indicate a 
coexistence of free and bound charges. The latters may be polaronic in nature
and may aggregate at low temperatures, as proposed 
for the NCCO system in Ref. \onlinecite{prl99}.
Emery and Kivelson\cite{Emery} have presented a picture where free carriers 
are scattered by arrays of (dynamical) bound charges which carry local 
dipoles. Following these authors, the optical conductivity can be written as 

\begin{equation}
\sigma (\omega,T) = \sigma_a + \sigma_b =
{{e^2A} \over \omega} \chi_2(\omega,T) + (e^*)^2 \omega 
\chi_2(\omega,T) 
\label{eq4}
\end{equation}

\noindent
with 

\begin{equation}
\chi_2(\omega,T) =  c \tanh(\hbar \omega/2k_BT) 
{\Gamma_p \over {\Gamma_p^2 + \omega^2}} \, .  
\label{eq5}
\end{equation}

\noindent
In Eq. (4), $\sigma_b$ is the
contribution of the bound charges, modeled as $c$ dipoles of charge $e^*$. 
$\sigma_a$ takes into account the scattering of the free carriers of charge 
$e$ by these dipoles. $\Gamma_p$ determines both the FIR peak frequency 
and
width, while $A$ is a constant whose  expression is reported in Ref. 
\onlinecite{Emery}. 

\begin{figure}
{\hbox{\psfig{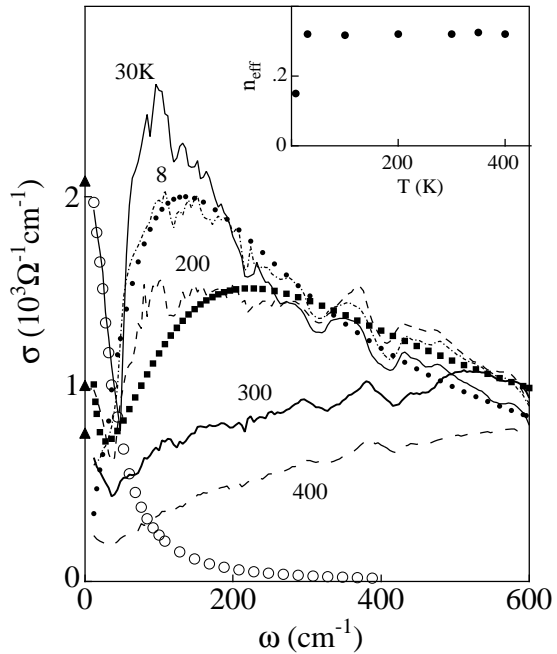}}}
\caption{Expansion in the FIR of the $\sigma (\omega)$ of Fig. 2. The 
open circles are the best fit to a normal Drude term, given by Eq. (3), here 
proposed only for 
the 30 K curve. The dots and the squares are best fits of Eqs. (4) and (5) to 
data at 8 K and 200 K, respectively. The triangles on the ordinate axis 
represent the values of
$\sigma_{dc}$ measured at the same temperatures, also  
reported in the inset of Fig. 1.}
\label{3}
\end{figure}

Equation (4) was compared unsuccessfully with the optical conductivity 
measured in La$_2$CuO$_{4+y}$.\cite{Tanner} On the contrary, good fits 
of Eq. (4) to the present BSCO data are obtained at all temperatures, 
as reported in Fig. 3 for $T > T_c$ (200 K) and $T < T_c$ (8 K). 
Remarkably, by using the two parameters $c$ and $\Gamma_p$ (whose 
values are 230 cm$^{-1}$ at 200 K, 135 cm$^{-1}$ at 8 K), one
reproduces the Drude term, the asymmetric FIR peak, and the softening 
and narrowing of this latter for decreasing $T$. 
 
Below $T_c$ the Drude peak disappears from the measuring range 
and the FIR peak looses the intensity corresponding to 
the underlying Drude tail. This effect, which 
has been reproduced in the fit by fixing 
$\sigma_a (\omega, T=8 K)$ = 0,  allows one to perform a further 
check of the present analysis. In fact, the loss of spectral weight observed 
below $T_c$ in Fig. 3 provides an estimate of the London penetration depth 
in the film, as given by the Ferrell-Glover sum rule\cite{Tinkham}

\begin{equation}
\lambda^2_L = { c^2 \over 8 \int _{\omega_{min}}^{\omega_{max}}
[ \sigma_n (\omega, 30 K) - \sigma_s (\omega, 8 K)]\, d \omega} \,,
\label{eq6}
\end{equation}

\noindent
where here $c$ is the speed of light and the other symbols have a obvious 
meaning. For $\omega_{min}$ = 10 
cm$^{-1}$ and $\omega_{max}$ = 15,000 
cm$^{-1}$ one obtains $\lambda_L = 300 \pm 10$ nm, a value in excellent 
agreement with that obtained for Bi$_{2}$Sr$_2$CuO$_{6}$ 
from transport measurements (310 nm).\cite{Martin}
On the other hand, no estimate of the superconducting gap can be made, 
due to the strong FIR peak which overshadows
the collapse of the Drude component  below $T_c$.

In conclusion, the present far-infrared study of BSCO questions the widely 
accepted ``anomalous Drude + pseudogap'' model,  through the identification 
in $\sigma (\omega,T)$ of {\it a narrow, normal Drude absorption well 
resolved 
from a broad FIR peak}. The small value of the free-carrier scattering 
rate ($\Gamma_D <$ 50 
cm$^{-1}$) is consistent with a metal in an extremely clean limit, where the 
carriers
move in well ordered, nearly defect-free Cu-O planes. Comparable values of 
$\Gamma_D$ can be obtained in doped semiconductors by the technique of 
''modulation doping,'' namely, by spatially separating the impurities, which
provide the free carriers, from the planes where these latters 
are free to move. In BSCO the Cu-O planes seem then to be very clean, while 
the eventual impurities and defects are mostly distributed out-of-plane.

The $\omega^{-1}$ dependence of $\sigma (\omega)$ in the mid infrared, 
often invoked to justify the need for an anomalous Drude model, is due 
to the tail of the 
peak centered at finite frequencies and should be ascribed to the bound 
charges. Similarly, the optical pseudogap (see Fig. 2) is a depression 
in $\sigma (\omega)$ created by
the bound-charge absorption as this latter narrows and 
softens for decreasing $T$. 
This effect is required by the conservation of $n_{eff}$ with temperature, 
here fulfilled within a few percent. In this respect, any optical determination 
of the temperature $T^*$ where the pseudogap starts opening seems 
questionable, 
as typically the FIR peak is observed even at $T>T^*$.

The present observations strongly 
support the coexistence of free and weakly bound charges in BSCO, 
responsible 
for the normal Drude absorption and the FIR peak, 
respectively. This latter is very similar to the polaronic feature observed in 
semiconducting cuprates at low 
doping. Its softening for decreasing $T$, followed by a saturation below $ T 
\sim$ 150 K, has already been attributed in Nd$_{2-x}$Ce$_x$CuO$_4$ to 
the formation of polaronic aggregates. This approach is 
confirmed by the present observations in 
BSCO, which are remarkably well fitted by a phase-separation model based 
on the scattering of the free carriers by  dynamical arrays of weakly bound 
charges. 

We thank G. Balestrino for providing the sample here investigated. We are 
also indebted to C. Di Castro, R. Gonnelli, and M. Grilli for useful 
discussions.

\end{multicols}

\end{document}